\documentclass[showpacs,epsf,onecolumn]{revtex4-1}
\usepackage{float}
\usepackage{color,graphicx}
\graphicspath{ {images/} }
\usepackage{hyperref}
\begin{document}

\title{Fermi surface topology and signature of surface Dirac nodes in LaBi}

\author{Ratnadwip Singha, Biswarup Satpati, Prabhat Mandal\footnote{email:prabhat.mandal@saha.ac.in}}

\affiliation{Saha Institute of Nuclear Physics, HBNI, 1/AF Bidhannagar, Calcutta 700 064, India}
\date{\today}

\begin{abstract}
Novel topological state of matter is one of the rapidly growing fields in condensed matter physics research in recent times. While these materials are fascinating from the aspect of fundamental physics of relativistic particles, their exotic transport properties are equally compelling due to the potential technological applications. Extreme magnetoresistance and ultrahigh carrier mobility are two such major hallmarks of topological materials and often used as primary criteria for identifying new compounds belonging to this class. Recently, LaBi has emerged as a new system, which exhibits the above mentioned properties. However, the topological nature of its band structure remains unresolved. Here, using the magnetotransport and magnetization measurements, we have probed the bulk and surface states of LaBi. Similar to earlier reports, extremely large magnetoresistance and high carrier mobility have been observed with compensated electron and hole density. The Fermi surface properties have been analyzed from both Shubnikov-de Haas and de Haas-van Alphen oscillation techniques. In the magnetization measurement, a prominent paramagnetic singularity has been observed, which demonstrates the non-trivial nature of the surface states in LaBi. Our study unambiguously confirms that LaBi is a three-dimensional topological insulator with possible linear dispersion in the gapped bulk band structure.
\end{abstract}

\maketitle

Materials with large magnetoresistance (MR) have always attracted attention due to their possible applications in magnetic valves, sensors and memory device technology$^{1-3}$. While the realization of giant and colossal MR have been proved to be major steps forward in this field$^{4-6}$, the recent discovery of extreme MR (XMR) in topological materials$^{7-9}$, has further stimulated interests in such systems. Here, the XMR is believed to appear, when the applied magnetic field lifts the symmetry protection of the spin-locked states$^{7}$. The coexistence of topologically distinct surface and bulk states is the unique characteristic feature of electronic band structure in topological insulators (TI) and semimetals (TSM). In TI, the time reversal symmetry protected metallic surface state is accompanied by insulating bulk state, whereas TSM hosts linear band crossings in their bulk band structure. Furthermore, the discovery of three-dimensional Dirac/Weyl fermions as quasi particle excitations in the bulk band structure$^{10,11}$ and observation of relativistic phenomena such as Adler-Bell-Jackiw chiral anomaly (signature of Weyl fermions)$^{12,13}$, have made the TSM as one of the fascinating topics in condensed matter physics research. Therefore, theoretical prediction and experimental realization of new materials belonging to different topological classes, are going on in full swing.

Recently, from the first-principles calculations$^{14}$, the members of the family of rare earth monopnictides La\textit{X} (\textit{X} = N, P, As, Sb, Bi) have been predicted to be potential candidates for TSM and TI. Several magnetotransport studies have demonstrated XMR in both LaSb and LaBi$^{15-18}$. However, little information is available on the topological nature of their band structure. In fact, for both the systems, the XMR has been attributed to the compensated electron and hole density rather than any non-trivial electronic band topology$^{16,19}$. On the other hand, it is always experimentally challenging to probe the topological nature of the surface state, which can settle the debate once and for all. LaBi is the end member of the La\textit{X} family with strongest spin-orbit coupling (SOC) and shares many similarities with LaSb. With increasing atomic number from N to Bi, the increasing SOC is predicted to open a gap in the bulk at the band crossing points and thus can induce a topological transition from semimetal to insulating phase$^{14}$. While contradicting ARPES reports exist in literature claiming both trivial and non-trivial topological states in LaSb$^{20,21}$, recent ARPES studies have demonstrated multiple Dirac nodes at the surface of LaBi$^{21-24}$. Therefore, detailed investigations are required, which would not only resolve the topological nature of LaBi but also can shed some light on the electronic band structure of other members of this family.

Herein, we have used the magnetotransport and magnetization measurements to probe the nature of the bulk and surface electronic band structure of LaBi. In several earlier studies on LaBi, the Fermi surface properties have been analyzed from the Shubnikov-de Haas (SdH) oscillation observed in the MR. Despite high sample quality, the number of Fermi pockets reported by different groups do not converge$^{16,17,25}$. This discrepancy in reported results may be the inherent problem of this technique. As the SdH oscillation is sensitive to quantum interference effects or noise from the electrical contacts, it may not be possible to separate some oscillation components, especially when the amplitude of oscillation is small. Therefore, to complement the SdH oscillation results we have also analyzed the de Haas-van Alphen (dHvA) oscillation in the magnetization measurement. This technique is not only free from interference effects, but the appearance of much sharper and readily distinguishable oscillation peaks down to very low magnetic field suggests that dHvA is a much better probe to study the Fermi surface properties than the SdH technique, which often requires high magnetic fields. The calculated Berry phase from both the oscillations confirms the presence of three-dimensional (3D) Dirac fermions in LaBi. In addition to the quantum oscillations, a robust paramagnetic singularity in magnetic susceptibility is observed at low fields, which originates from the electronic states near 2D surface Dirac nodes of a TI. Thus the present work reveals the topological insulating phase in LaBi together with linear dispersion in the gapped bulk states. To the best of our knowledge, this is the first magnetization report on LaBi, which explores the topological nature of its band structure and confirms the theoretical prediction as well as the recent ARPES results. The non-trivial surface state of a topological material can also be explored through Aharonov-Bohm oscillation. However, this technique requires large surface-to-volume ratio to minimize the bulk dominated transport and hence can only be observed in samples with reduced dimension such as nanowire, nanoribbon etc$^{26,27}$.\\

\textbf{Results}\\

\textbf{Sample characterization.} Fig. 1(a) shows the X-ray diffraction (XRD) pattern for the LaBi single crystal along the (1 0 0) plane. Presence of very sharp (h 0 0) peaks in diffraction pattern confirm the high crystalline nature of the grown crystals. LaBi crystallizes in a rock salt type structure with space group \textit{Fm$\bar{3}$m}. From the XRD pattern, the lattice parameter is deduced to be \textbf{a}=6.5703(2) ${\AA}$, which is consistent with the earlier reports$^{17,25}$. In the inset, a typical single crystal is shown with different crystallographic directions, along which measurements have been performed. The \textbf{a}-axis is along [h 0 0] direction. The XRD has been performed on several single crystals from the same batch. While the peak intensity has been seen to vary, no impurity phase was detected within the experimental resolution. In Fig. 1(b), the high-resolution transmission electron microscopy (HRTEM) image clearly illustrates the high quality crystalline nature of the LaBi samples. From the Fourier filtered HRTEM image [Fig. 1(c)], an interlayer spacing d=3.32(4) ${\AA}$ is determined along (2 0 0) plane, which agrees with that calculated from the XRD data (d=a/2). The energy-dispersive X-ray (EDX) spectroscopy data, shown in Fig. 1(d), confirm almost perfect stoichiometry [La:Bi = 1:1.08] and the absence of any impurity in the LaBi crystals. The maximum relative error in the calculated atomic ratio is $\sim$5\%. The copper and carbon signals in the spectrum are coming from the carbon coated copper grid on which the samples were mounted for HRTEM.\\

\begin{figure}
\includegraphics[width=0.6\textwidth]{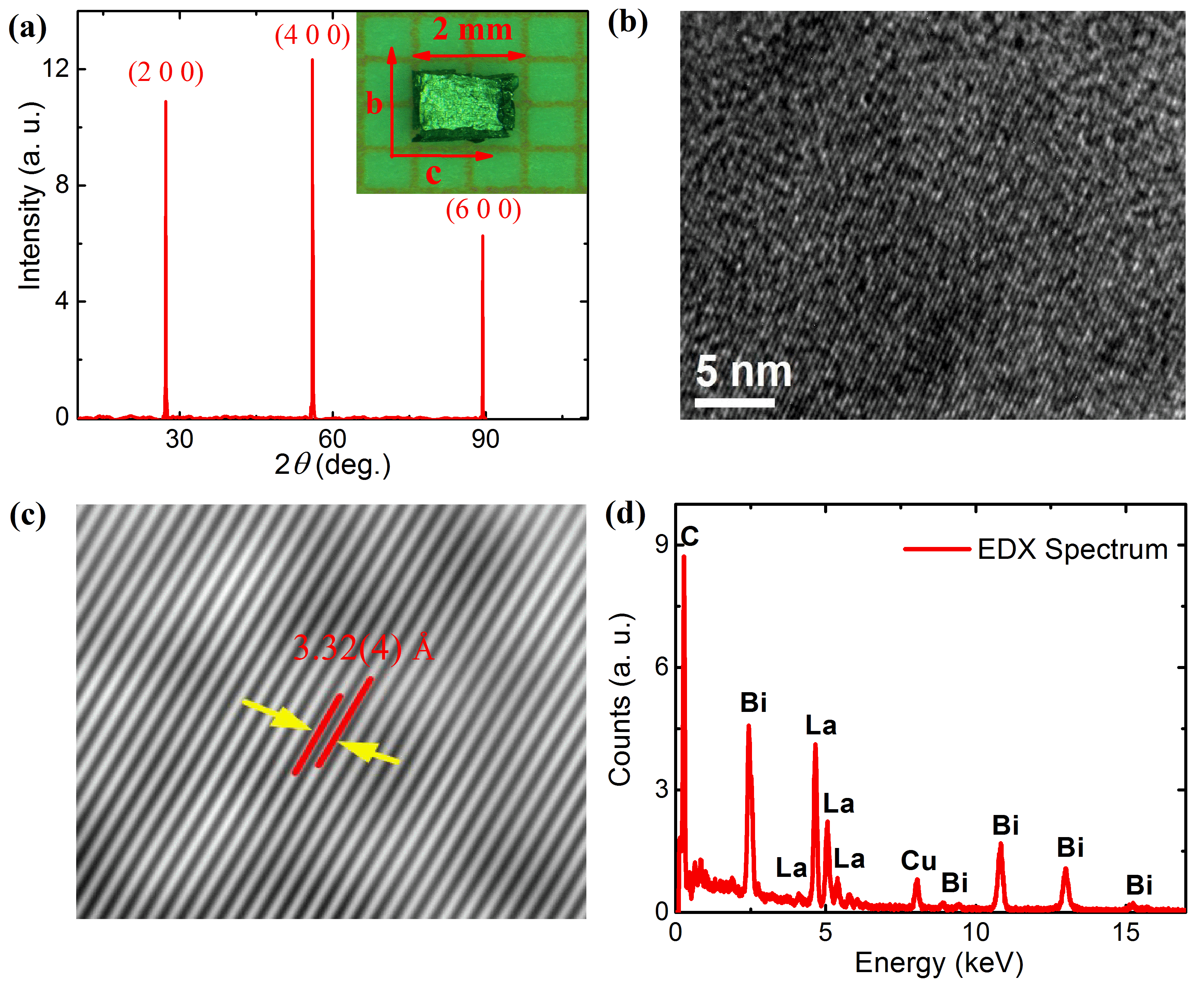}
\renewcommand{\figurename}{\textbf{Figure}}
\caption{\textbf{Characterization of as grown LaBi single crystal.} (a) XRD pattern of the measured samples. Inset shows a typical single crystal with different crystallographic directions. (b) HRTEM image of the grown crystal. (c) Fourier filtered HRTEM image showing the interlayer spacing along (2 0 0) plane. (d) The EDX spectroscopy data of the grown crystals. The copper and carbon signals are coming from the carbon coated copper grid on which the samples were mounted.}
\end{figure}

\textbf{Temperature dependence of resistivity and extreme magnetoresistance.} The resistivity of LaBi single crystal has been measured with current along \textbf{c}-axis and magnetic field along \textbf{a}-axis. As shown in Fig. 2(a), the zero-field resistivity ($\rho_{xx}$) is metallic in nature and decreases monotonically as the temperature decreases. At 2 K, $\rho_{xx}$ becomes as small as $\sim$130 $n\Omega$ cm, yielding a large residual resistivity ratio [RRR = $\rho_{xx}$(300 K)/$\rho_{xx}$(2 K)] $\sim$350. At low temperature the resistivity obeys a power-law behavior, $\rho_{xx}(T)$ = $\rho_{0} + AT^{n}$ with $n$$\sim$3. When the magnetic field is applied, $\rho_{xx}$ is seen to increase and a metal-semiconductor like crossover starts to appear along with resistivity plateau. Such kind of resistivity behavior is a generic feature for the TSMs$^{8,9,28}$ and can be explained from Kohler scaling analysis$^{29,30}$. From the $\partial\rho_{xx}/\partial T$ curves [Fig. 2(a) inset], two characteristic temperatures can be clearly identified. The crossover temperature ($T_{m}$), where the slope of $\rho_{xx}(T)$ curve changes its sign and $T_{i}$, the point of inflection. Slightly below this temperature ($T_{i}$), resistivity starts to saturate. $T_{m}$ increases monotonically with field and follows a $T_{m}\propto(B-B_{0})^{1/\nu}$ type relation with $\nu\sim$2 above a critical field $B_{0}$$\sim$1 T, which is similar to that observed in several compensated semimetals $^{29,31}$. On the other hand, $T_{i}$ is almost independent of the applied magnetic field strength.

With the same current and magnetic field configurations, an extremely large and non-saturating MR $\sim$ 4.4$\times$10$^{4}$ \% is obtained at 2 K and 9 T [Fig. 2(b)], which is consistent with the earlier reports on LaBi$^{16,17}$ and comparable to that reported for several TSM candidates$^{29,32}$. However, with increasing temperature the value of MR decreases rapidly and becomes only $\sim$ 6 \% at 300 K and 9 T. The fitting in the inset of Fig. 2(b) suggests that the MR curve obeys a power law, MR $\propto B^{m}$ with $m\sim$1.7. The observed power-law behavior is close to the parabolic field dependence (MR$\propto$$B^{2}$), which is expected for compensated semimetallic systems from classical two-band theory$^{33}$.\\

\begin{figure}
\includegraphics[width=0.6\textwidth]{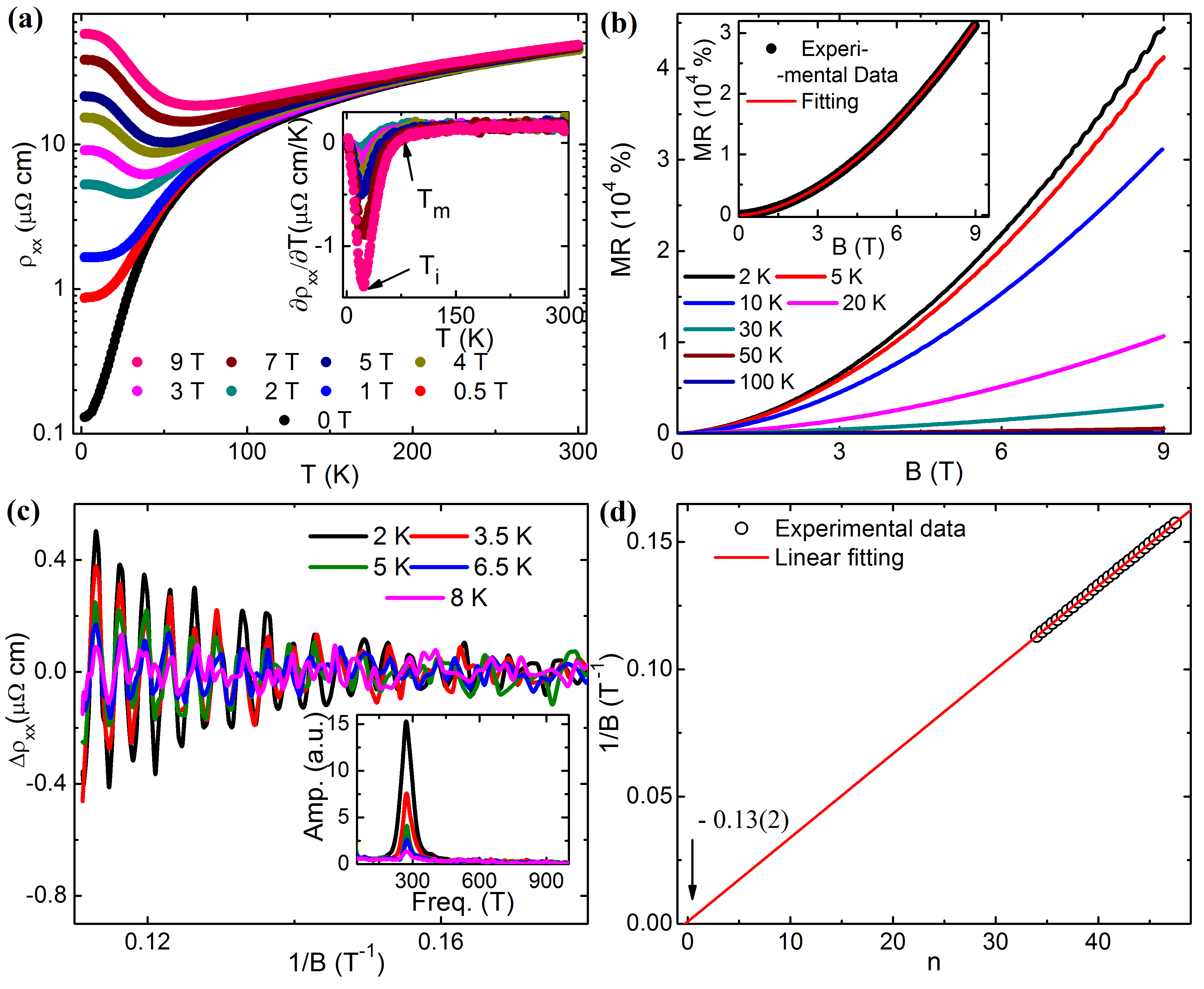}
\renewcommand{\figurename}{\textbf{Figure}}
\caption{\textbf{Temperature dependence of resistivity and transverse MR in LaBi single crystals.} (a) Temperature dependence of resistivity at different magnetic field strengths. Current and magnetic field are along \textbf{c} and \textbf{a}-axis, respectively. Inset shows the temperature dependence of first-order derivative of resistivity ($\partial\rho_{xx}/\partial T$) at different magnetic fields with characteristic temperatures $T_{m}$ and $T_{i}$. (b) Transverse MR with current along \textbf{c} and magnetic field parallel to \textbf{a} axis, at several representative temperatures. Inset illustrates the power-law behavior of the observed MR at 10 K. (c) SdH oscillation obtained by subtracting polynomial background from MR measurement, plotted with inverse magnetic field (1/B) at different temperatures. Inset shows the corresponding FFT result. (d) Landau level index plot obtained from SdH oscillation. The value of x-axis intercept is shown by the arrow.}
\end{figure}

\begin{figure}
\includegraphics[width=0.35\textwidth]{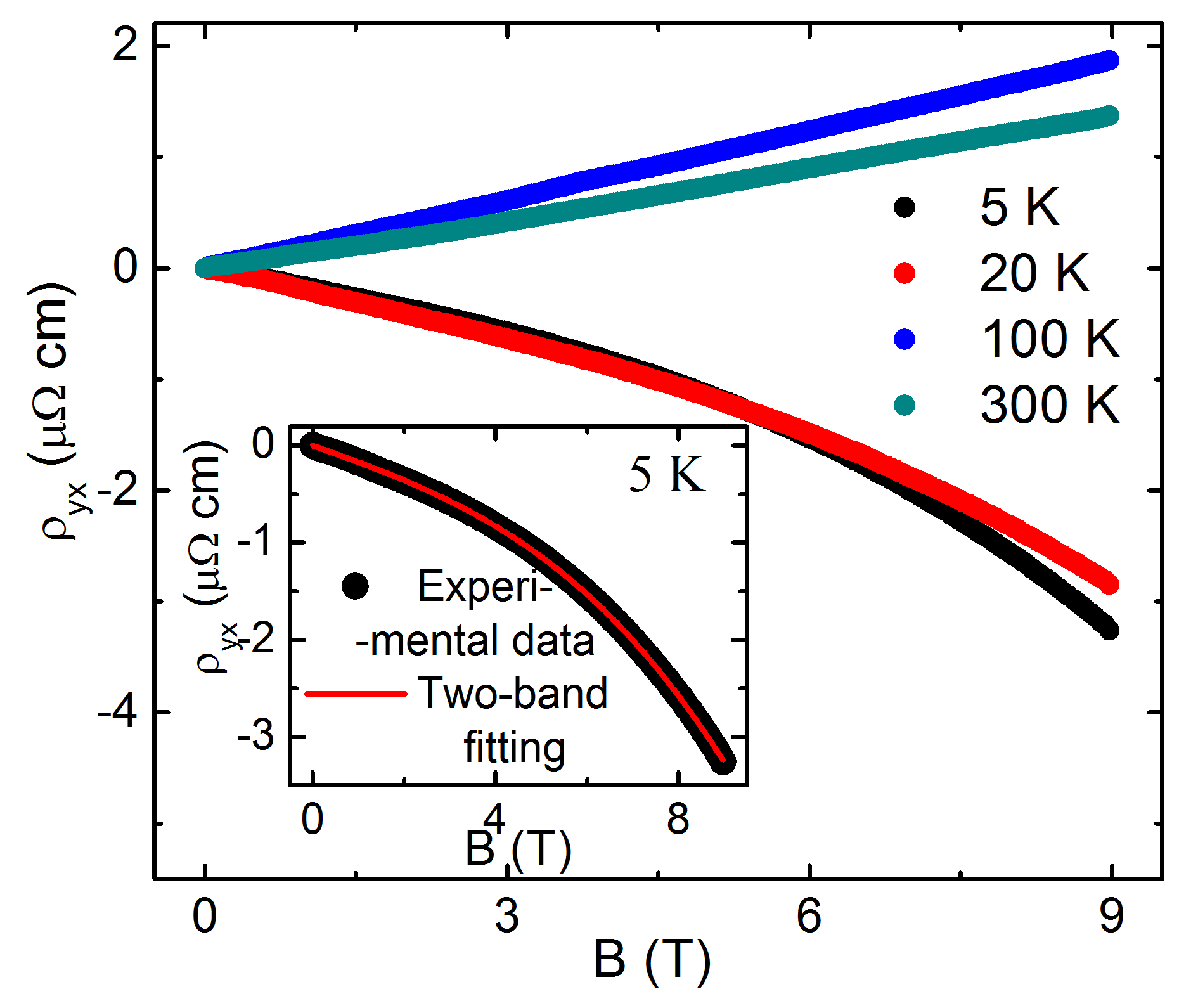}
\renewcommand{\figurename}{\textbf{Figure}}
\caption{\textbf{Hall resistivity for LaBi single crystal.} Magnetic field dependence of Hall resistivity at different temperatures. Inset shows the classical two-band fitting of $\rho_{yx}$.}
\end{figure}

\textbf{Shubnikov-de Haas oscillation and Hall measurement.} In Fig. 2(c), the oscillatory component of the resistivity [$\Delta\rho_{xx}(B)$] is plotted after subtracting a fifth order polynomial background from the experimental data. To decide the appropriate polynomial order, we have subtracted higher as well as lower order polynomial background. However, no visible change has been noticed above fifth order, which replicate the background most appropriately. Also, by comparing the results obtained after polynomial background and smooth background subtraction, we did not find any difference. The fast Fourier transform (FFT) analysis of $\Delta\rho_{xx}(B)$ in the inset of Fig. 2(c), reveals an oscillation frequency 273(5) T, which is consistent with the earlier report by Tafti \textit{et al.}$^{25}$ On the other hand, both Sun \textit{et al.}$^{16}$ and Kumar \textit{et al.}$^{17}$ have reported the presence of an additional higher frequency component in LaBi. However, from our SdH oscillation data, we are not able to resolve any additional frequency, which may be due to very weak nature of oscillation for the higher frequency component within the measured field and temperature range. On the other hand, we have clearly resolved two Fermi pockets on the same piece of sample from our de Haas-van Alphen (dHvA) oscillation data to be discussed later. Using the Onsager relation, $F = (\phi_{0}/2\pi^{2})A_{F}$, where $\phi_{0}$ is the single magnetic flux quantum, we have calculated the Fermi surface cross-section ($A_{F}$) $\sim$2.6(1)$\times$10$^{-2}$ ${\AA}^{-2}$ perpendicular to the \textbf{a}-axis. The corresponding Fermi momentum ($k_{F}$) and Fermi velocity ($v_{F}$) are calculated to be 9.1(2)$\times$10$^{-2}$ ${\AA}^{-1}$ and 5.6(1)$\times$10$^{5}$ m/s, respectively. The oscillation amplitude can be fitted using the thermal damping factor of Lifshitz-Kosevich (L-K) formula, $R_{T} = (2\pi^{2}k_{B}T/\beta)/sinh(2\pi^{2}k_{B}T/\beta)$, where $\beta = e\hbar B/m^{\ast}$. From the fitting parameters, we have calculated the cyclotron mass ($m^{\ast}$) $\sim$ 0.19(1)$m_{0}$ for the charge carriers, where $m_{0}$ is the free electron rest mass. The field-induced damping of the oscillation amplitude can be fitted well with $R_{D} = exp(-2\pi^{2}k_{B}m^{\ast}T_{D}/\hbar eB)$ and the Dingle temperature ($T_{D}$) is determined to be 11.3(2) K. Using this Dingle temperature, we have also calculated the quantum mobility $\mu_{q}$ [= $(e\hbar/2\pi k_{B}m^{\ast}T_{D})$] $\sim$ 10$^{3}$ cm$^{2}$ V$^{-1}$ s$^{-1}$, which can provide a rough estimate on the mobility of the charge carriers. However, the quantum mobility is always expected to be lower than the classical Drude mobility$^{34}$.

The Berry phase acquired by the charge carriers can provide more information about the nature of electronic band structure in LaBi. According to Lifshitz-Onsager rule, a closed orbit in magnetic field is quantized as$^{35}$,
\begin{equation}
A_{F}\frac{\hbar}{eB} = 2\pi(n+\frac{1}{2}-\beta-\delta) = 2\pi(n+\gamma-\delta),
\end{equation}
where $2\pi\beta$ is the Berry phase. $\delta$ is a phase shift, which has a value 0 and $\pm$ 1/8 for 2D and 3D band structures, respectively. The Berry phase is 0 for the conventional metals with parabolic band dispersion and $\pi$ for the Dirac/Weyl type electronic system with linear band dispersion. For quantum oscillation with single frequency, the parameter $\gamma-\delta = \frac{1}{2}-\beta-\delta$, can be extracted from the x-axis intercept in the Landau level index plot and will remain within the range -1/8 to +1/8 for 3D Dirac fermions$^{35}$. In Fig. 2(d), the Landau level fan diagram is plotted from the SdH oscillation, assigning maxima as integers and minima positions as half-integers. The obtained intercept [$\sim$-0.13(2)] is close to the theoretical range for 3D Dirac fermions but far from that expected for a parabolic dispersion (intercept $\sim$0.5), which clearly demonstrates the presence of 3D Dirac fermions in LaBi.

The density and classical Drude mobility of the charge carriers can be determined from the Hall resistivity data, as has been shown in Fig. 4. While the Hall resistivity ($\rho_{xy}$) is positive and almost linear at high temperature, it becomes slightly non-linear and changes sign at low temperature. Therefore, both electron and hole pockets are present in LaBi. In the inset of Fig. 3, the Hall resistivity is fitted using classical two-band model$^{36}$,
\begin{equation}
\rho_{xy} = \frac{B}{|e|}\frac{(n_{h}\mu_{h}^{2}-n_{e}\mu_{e}^{2})+(n_{h}-n_{e})(\mu_{h}\mu{e})^{2}B^{2}}{(n_{h}\mu_{h}+n_{e}\mu_{e})^{2}+(n_{h}-n_{e})^{2}(\mu_{h}\mu{e})^{2}B^{2}},
\end{equation}
where $n_{h}$ ($n_{e}$) and $\mu_{h}$ ($\mu_{e}$) are density and mobility of the holes (electrons), respectively. The obtained electron and hole densities $\sim$ 2$\times$10$^{19}$ cm$^{-3}$ and $\sim$ 1.9$\times$10$^{19}$ cm$^{-3}$, respectively, reveal that LaBi is a compensated semimetal. From the fitting parameters, the mobilities of the carriers are determined to be $\sim$1.28$\times$10$^{4}$ cm$^{2}$ V$^{-1}$ s$^{-1}$ and $\sim$1.26$\times$10$^{4}$ cm$^{2}$ V$^{-1}$ s$^{-1}$ for electrons and holes, respectively. The obtained carrier mobility is quite large and comparable to several topological Dirac and Weyl semimetals$^{28,34,37}$.\\

\begin{figure}
\includegraphics[width=0.6\textwidth]{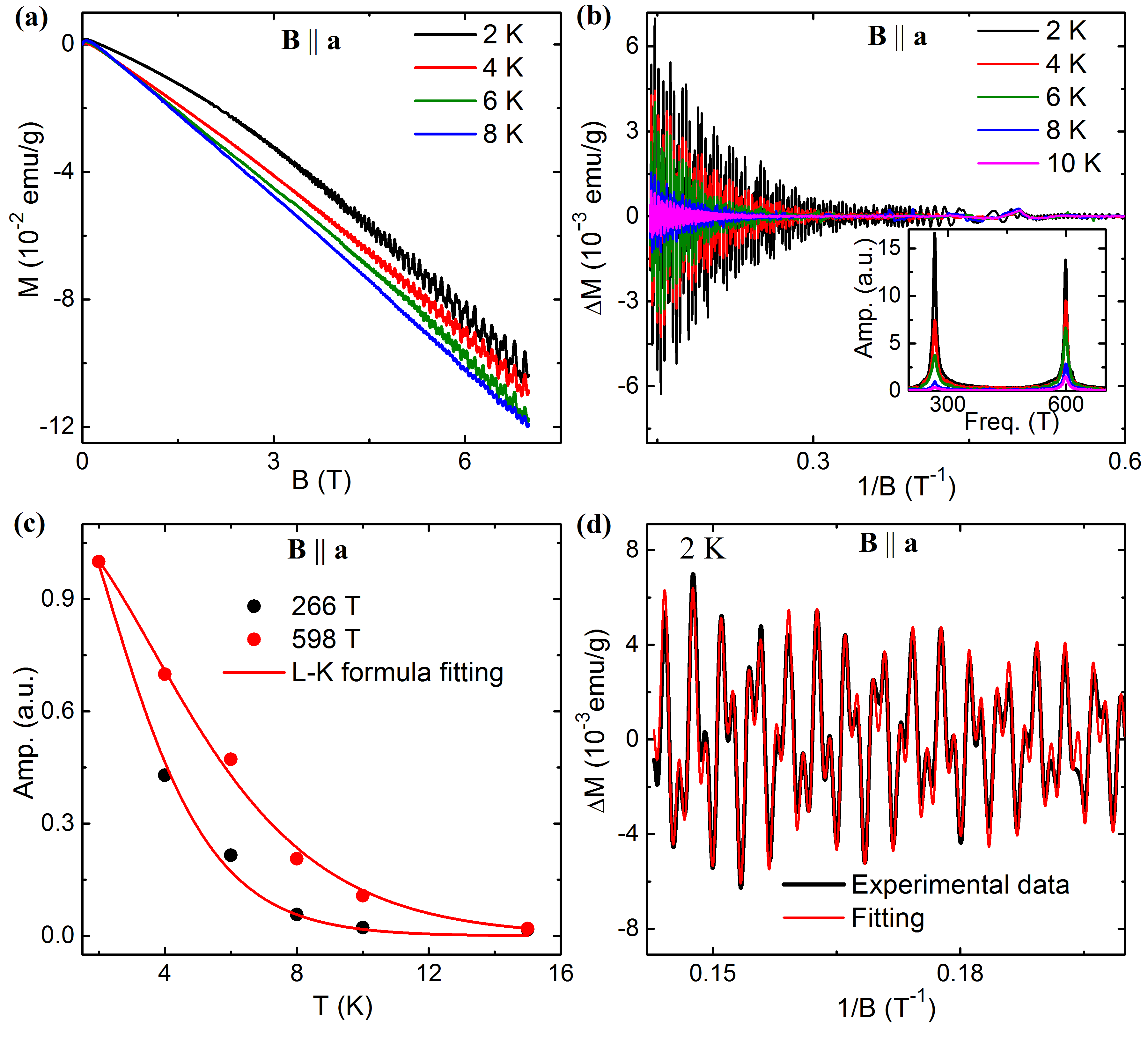}
\renewcommand{\figurename}{\textbf{Figure}}
\caption{\textbf{dHvA oscillation in LaBi single crystal.} (a) Magnetic moment plotted as a function of magnetic field with field parallel to \textbf{a}-axis. (b) dHvA oscillation obtained after background subtraction. Inset shows the corresponding FFT result. (c) Temperature dependence of dHvA oscillation amplitude, fitted using the L-K formula for \textbf{B}$\parallel$\textbf{a}-axis. (d) Two-band L-K formula fitting of dHvA oscillation.}
\end{figure}

\textbf{de Haas-van Alphen oscillation and Fermi surface properties.} In Fig. 4(a), the results of magnetization measurements have been shown with magnetic field along \textbf{a}-axis. Similar to other TSMs, a diamagnetic behavior has been observed along with very clear dHvA oscillations. In contrast to the SdH oscillation, which can be seen only up to 8 K, dHvA can be easily resolved even at 15 K. At lowest measured temperature (2 K), weak but detectable oscillations have been observed down to 0.8 T, which corresponds to a magnetic length $l_{B} = \sqrt{\hbar/eB}$ $\approx$ 29 nm. This large magnetic length indicates high quality of the LaBi single crystals. The oscillatory component ($\Delta$M) has been extracted after subtracting a smooth background and plotted in Fig. 4(b). In the inset, the FFT spectrum is shown, which clearly reveals two oscillation frequencies 266(1) T and 598(2) T. The presence of more than one frequency component is also evident from the small but sharp oscillation peak adjacent to each higher amplitude peak in Fig. 4(b) [also see Fig. 4(d)]. The sharp peaks in the FFT spectrum indicate almost no error in the frequency deduced using this techniques. On the other hand, due to the relatively broad nature of the peak in FFT spectrum, the error in the frequency value is higher in SdH oscillation compared to dHvA. This is expected as larger number of oscillation peaks have been used for dHvA oscillation analysis compared to few cycles of oscillation obtained in SdH measurement. Furthermore, quantum oscillations are expected to appear in very clean samples, i.e., where the conductivity and hence mean free path is very high. Therefore, to obtain appropriate electrical signals for SdH oscillation, quite high current is needed. This can cause local heating and blur the finer details of the oscillation. However, dHvA technique is free from such drawback. From the Onsager relationship, the Fermi surface cross-sections have been calculated to be 2.53(1)$\times$10$^{-2}$ ${\AA}^{-2}$ and 5.70(2)$\times$10$^{-2}$ ${\AA}^{-2}$, for smaller and larger pockets, respectively. The temperature damping of the FFT amplitude is shown in Fig. 4(c) and has been fitted using L-K formula. From the fitting parameters, we have deduced effective mass 0.11(3)$m_{0}$ and 0.07(2)$m_{0}$ for 266(1) T and 598(2) T, respectively. Prominent dHvA oscillations have also been observed, when the magnetic field is applied along other two crystallographic axes. While the oscillation frequencies are same for \textbf{a}- and \textbf{c}-axis, both peaks in FFT spectrum slightly shift towards higher values [272(1) T and 602(2) T] for magnetic field along \textbf{b}-axis. As LaBi possesses cubic symmetry, which has also been confirmed from our X-ray diffraction and transmission electron microscopy measurements, it is expected that these three crystallographic axis should be equivalent. On the other hand, the hole pocket in LaBi has been seen to be highly anisotropic$^{17}$. Therefore, a slight misalignment of the magnetic field and b-axis may result in this small deviation. The Fermi surface parameters calculated from both the quantum oscillation techniques are summarized in TABLE I. The small difference in the frequency values, obtained from these two techniques, may be partially due to a small misalignment of the magnetic field and or the higher error in the SdH frequency as we have discussed earlier. However, the difference in the number of Fermi pockets, probed in these two methods, can not be explained from such arguments. There are several reports on cyclotron resonance experiment and theoretical works for LaBi and isostructural compound LaSb, as these are quite old systems$^{38-41}$. Moreover, dHvA oscillation has been studied in LaSb$^{42}$. The cyclotron mass (varies between 0.13-0.33$m_{0}$ for two branches) and extremal Fermi surface cross-section calculated in those studies along [1 0 0] direction are in agreement with that obtained from our quantum oscillation data.

\begin{center}
\textbf{TABLE I: Fermi surface parameters extracted from SdH and dHvA oscillations.}\\
\end{center}

\begin{center}
 \begin{tabular}{||c c c c c c c||}
 \hline
  & Configuration & F & $A_{F}$ & $k_{F}$ & $m^{\ast}$ & $v_{F}$\\

  &  & T & 10$^{-2} {\AA}^{-2}$ & 10$^{-2} {\AA}^{-1}$ & $m_{0}$ & 10$^{5}$ m/s\\ [0.5ex]
 \hline\hline
 SdH & \textbf{B}$\parallel$\textbf{a} & 273(5) & 2.6(1) & 9.1(2) & 0.19(1) & 5.6(1)\\[0.5ex]

 \hline\hline

 dHvA & \textbf{B}$\parallel$\textbf{a} & 266(1) & 2.53(1) & 8.97(2) & 0.11(3) & 9.3(5)\\

  &  & 598(2) & 5.70(2) & 13.47(2) & 0.07(2) & 22(3)\\ [1ex]
 \hline
\end{tabular}
\end{center}

As two frequencies are involved in the dHvA oscillation in LaBi, the Berry phase can not be extracted from the simple Landau level index plot. In such case, the experimental data can be described as a superposition of damped cosine-function for each frequency component, which is given by the complete L-K formula$^{43,44}$,
\begin{equation}
\Delta M \propto -B^{1/2}R_{T}R_{D}R_{S} sin\left[2\pi\left(\frac{F}{B}+\gamma-\delta\right)\right].
\end{equation}
Here, $R_{T}$ and $R_{D}$ are the thermal and field damping factors, respectively, as discussed before and $R_{S} = cos(\pi gm^{\ast}/m_{0})$. Fig. 4(d) shows the two-band L-K formula fitting with the experimental data. From the fitting parameters, the Berry phase has been calculated to be 2$\pi$(0.50(3)+$\delta$) and 2$\pi$(0.46(2)+$\delta$) for 266(1) T and 598(2) T frequency branches, respectively, which are close to the $\pi$ Berry phase expected for the Dirac fermions. Small deviation in Berry phase value is quite common in these types of materials and can be associated with the Fermi surface anisotropy$^{44,45}$. The obtained Berry phase for both the Fermi pockets confirms the non-trivial nature of the electronic band structure in LaBi.

\begin{figure}
\includegraphics[width=0.6\textwidth]{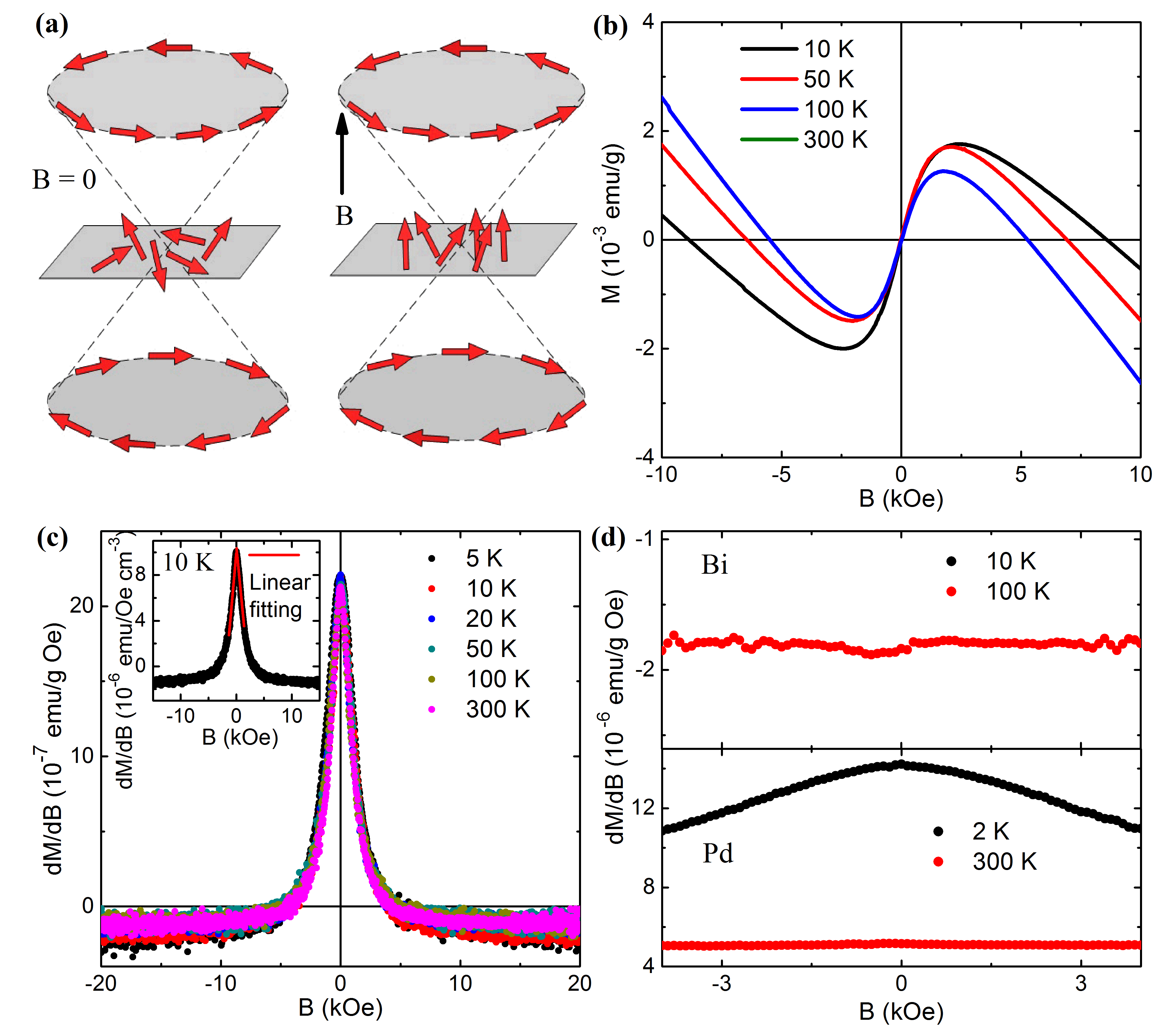}
\renewcommand{\figurename}{\textbf{Figure}}
\caption{\textbf{Paramagnetic singularity in magnetization measurement for LaBi.} (a) Schematic representing the helical spin-texture of surface Dirac cones for a 3D topological insulator without and with magnetic field. The arrows indicate the direction of the electron spins. (b) Low field region of the magnetic moment vs. $B$ curves at different temperatures for LaBi. (c) Magnetic susceptibility $\chi$ ($=dM/dB$), calculated by taking first order derivative of magnetic moment, plotted as a function of magnetic field. Inset shows the linear field decay of $\chi$. (d) Magnetic susceptibility of standard Bi and Pd samples.}
\end{figure}

\textbf{Paramagnetic singularity in magnetization measurement.} The main characteristic feature of a 3D TI is the gapless surface state with odd number of Dirac nodes, which are protected by time reversal symmetry. The low-energy physics of these metallic surface states can be mathematically formulated using a Dirac-type effective Hamiltonian$^{46}$, $H_{surf}(k_{x}, k_{y}) = \hbar v_{F}(\sigma^{x}k_{y}-\sigma^{y}k_{x})$, which leads to phenomenon like spin-momentum locking. Here, $\vec{\sigma}$ is the Pauli matrix. Hence, for a fixed translational momentum \textbf{k}, the spin has a fixed direction parallel to the sample surface$^{46,47}$. A corresponding helicity operator can be defined$^{46}$, $\hat{h} = (1/k)\hat{z}.(\vec{k}\times\vec{\sigma})$, which yields values +1 and -1 for the lower and upper Dirac cones of the surface band structure, respectively. Therefore, each Dirac cone possesses a particular helicity - upper Dirac cone with left-handed and lower Dirac cone with right-handed spin texture. On the other hand, the electron spins at the Dirac point do not have any preferable orientation and are free to align along the external magnetic field as long as the Dirac cone is not gapped. In the low-field region, these freely oriented spins are predicted to generate a paramagnetic singularity in the $M(B)$ curve, which should be reflected as a cusp in $\chi(B)$ plot$^{48}$. The spin-texture of the surface state for a 3D topological insulator is shown schematically in Fig. 5(a) both without and with magnetic field. Fig. 5(b) shows the low-field region of the $M(B)$ curves for LaBi, when the magnetic field is applied along \textbf{c}-axis. While the magnetization data predominantly demonstrate diamagnetic behavior throughout the measured field range, a clear paramagnetic contribution is observed in the vicinity of zero field. In Fig. 5(c), the corresponding susceptibility plots are shown at different representative temperatures. As predicted theoretically$^{48}$, robust cusps have been observed at $B$=0, which persist up to room temperature. Similar low-field paramagnetic singularity, which originates due to the topologically non-trivial surface electronic states, has been previously reported in 3D TIs such as Bi$_{2}$Se$_{3}$, Bi$_{2}$Te$_{3}$ and Sb$_{2}$Te$_{3}$, Bi$_{1.5}$Sb$_{0.5}$Te$_{1.7}$Se$_{1.3}$ and narrow gap topological insulator ZrTe$_{5}$ with gapped bulk Dirac cones$^{48-51}$. With both temperature and chemical potential set to zero, the susceptibility is given by$^{48}$,
\begin{equation}
\chi(B) \cong \chi_{0} + \frac{\mu_{0}}{4\pi^{2}}\frac{x}{L}\left[\frac{(g\mu_{B})^{2}}{\hbar v_{F}}\Lambda - \frac{2(g\mu_{B})^{3}}{\hbar^{2}v_{F}^{2}}\mid B\mid\right],
\end{equation}
where $\chi_{0}$, $\Lambda$, $\mu_{B}$ and $g$ are the background contribution, effective size of the momentum space contributing to the singular part of the total free energy, Bohr magneton and Land$\acute{e}$ $g$-factor, respectively. $L$ is the thickness of the measured crystal and $x$ is the fraction of the surface contributing, which is usually extremely small. As Eq. 4 suggests, $\chi(B)$ should decrease linearly with field, which is indeed observed in LaBi [Fig. 5(c) inset]. The measurements were performed on several crystals from the same batch. While the cusp width varies slightly with the dimensions, as expected from Eq. 4, the general behavior remains unaltered. The results of the measurement on standard diamagnetic (Bi) and paramagnetic (Pd) samples [Fig. 5(d)] do not show any low-field singularity. Moreover, such a unique temperature independent behavior of $\chi$ for LaBi can not be explained assuming a small magnetic impurity in diamagnetic host and is distinct from the magnetic response of dilute magnetic semiconductors. So, the low-field magnetic response must be a intrinsic property of the measured crystals. Therefore, LaBi certainly hosts topological Dirac fermions in the surface state, which is in accordance with the recent ARPES reports$^{21-24}$.\\

\textbf{Discussion and Conclusions}\\

LaBi has emerged lately as another new system with XMR and large carrier mobility. While a number of studies have been focused on the magnetotransport properties of LaBi, the topology of the electronic band structure remains substantially unexplored. Furthermore, the XMR is believed to appear in the electron-hole resonance regime due to the compensated semimetallic nature of LaBi. On the other hand, from the recent ARPES measurements, multiple Dirac cones have been observed at the surface of LaBi, which support the picture of a topological insulator, as also predicted by band structure calculation. In this report, we have combined the magnetotransport and magnetization measurements to probe the bulk Fermi surface along with the surface electronic states of LaBi. Similar to earlier reports, a non-saturating XMR ($\sim$ 4.4$\times$10$^{4}$ \%) and field-induced resistivity turn-on behavior have been observed. The measured non-linear Hall resistivity reveals the presence of two types of carriers with high carrier mobility and almost perfectly compensated densities. The Fermi surface parameters have been calculated from both SdH and dHvA oscillations. Although both the techniques are widely used to study Fermi surface properties, our results clearly show the advantages of dHvA technique over the other. In the low-field region of the magnetization, for the first time in LaBi, a robust paramagnetic singularity has been detected, which persists up to room temperature. This behavior of magnetic susceptibility is the signature of the helical spin texture of a 3D TI surface state. On the other hand, the calculated Berry phase from both SdH and dHvA oscillations confirms the 3D Dirac fermionic nature of the charge carriers. These results can only be explained assuming a small gap in the bulk band structure due to the strong SOC and linearly dispersing bulk bands. Therefore, LaBi must be a three-dimensional topological insulator, which also hosts gapped Dirac cone in its bulk state. Thus from the present work we can comprehensively conclude the debate over the topological nature of the band structure in LaBi and confirm the results of ARPES and band structure calculations.\\

\textbf{Methods}

High quality single crystals of LaBi were grown using indium flux$^{16,52}$. Elemental La (Alfa Aesar 99.9\%), Bi (Alfa Aesar 99.999\%) and In (Alfa Aesar 99.999\%) were taken in an alumina crucible in the molar ratio 1:1:20. The crucible was then sealed in a quartz tube under vacuum. The quartz tube was heated to 1000$^{\circ}$C, kept at this temperature for 5 h and then cooled slowly (4$^{\circ}$C/h) to 700$^{\circ}$C. At this temperature, indium is decanted in a centrifuge. Several single crystals of typical dimension 2$\times$1.8$\times$1 mm were obtained. The X-ray diffraction (XRD) spectrum of the single crystals was obtained in a Rigaku X-ray diffractometer. HRTEM of the grown crystals was done in an FEI, TECNAI G$^{2}$ F30, S-TWIN microscope operating at 300 kV and equipped with a GATAN Orius SC1000B CCD camera. EDX spectroscopy was performed using the same microscope with a high-angle annular dark-field scanning (HAADF) detector from Fischione (Model 3000). Transport measurements were performed in a 9 T Physical Property Measurement System (Quantum Design) by four-probe technique using the ac transport option. Magnetic measurements were done in a 7 T SQUID-VSM MPMS 3 (Quantum Design). Before doing the magnetic measurements on LaBi, empty sample holders were measured to ensure the absence of any contamination. The obtained data are more than two orders of magnitude smaller than that obtained with LaBi samples. The measurements were performed on several single crystals from the same batch, all producing similar results.\\

\textbf{References}\\

1. Daughton, J. M. GMR applications. \textit{J. Mag. Mag. Mater.} \textbf{192}, 334-342 (1999).

2. Wolf, S. A. \textit{et al.} Spintronics: A spin-based electronics vision for the future. \textit{Science} \textbf{294}, 1488-1495 (2001).

3. Lenz J. A review of magnetic sensors. \textit{Proc. IEEE} \textbf{78}, 973-989 (1990).

4. Baibich, M. N. \textit{et al.} Giant Magnetoresistance of (001)Fe/(001)Cr Magnetic Superlattices. \textit{Phys. Rev. Lett.} \textbf{61}, 2472-2475 (1988).

5. Binash, G., Gr\"{u}nberg, P., Saurenbach, F. \& Zinn, W. Enhanced magnetoresistance in layered magnetic structures with antiferromagnetic interlayer exchange. \textit{Phys. Rev. B} \textbf{39}, 4828-4830 (1989).

6. Tokura, Y. Critical features of colossal magnetoresistive manganites. \textit{Rep. Prog. Phys.} \textbf{69}, 797-851 (2006).

7. Liang, T. \textit{et al.} Ultrahigh mobility and giant magnetoresistance in the Dirac semimetal Cd$_{3}$As$_{2}$. \textit{Nat. Mater.} \textbf{14}, 280-284 (2015).

8. Ali, M. N. \textit{et al.} Large, non-saturating magnetoresistance in WTe$_{2}$. \textit{Nature} \textbf{514}, 205-210 (2014).

9. Shekhar, C. \textit{et al.} Extremely large magnetoresistance and ultrahigh mobility in the topological Weyl semimetal candidate NbP. \textit{Nat. Phys.} \textbf{11}, 645-649 (2015).

10. Liu, Z. K. \textit{et al.} A stable three-dimensional topological Dirac semimetal Cd$_{3}$As$_{2}$. \textit{Nat. Mater.} \textbf{13}, 677-681 (2014).

11. Xu, S. Y. \textit{et al.} Discovery of a Weyl fermion semimetal and topological Fermi arcs. \textit{Science} \textbf{349}, 613-617 (2015).

12. Huang, X. \textit{et al.} Observation of the Chiral-Anomaly-Induced Negative Magnetoresistance in 3D Weyl Semimetal TaAs. \textit{Phys. Rev. X} \textbf{5}, 031023 (2015).

13. Li, C.-Z. \textit{et al.} Giant negative magnetoresistance induced by the chiral anomaly in individual Cd$_{3}$As$_{2}$ nanowires. \textit{Nat. Comm.} \textbf{6}, 10137 (2015).

14. Zeng, M. \textit{et al.} Topological semimetals and topological insulators in rare earth monopnictides. \textit{arXiv:1504.03492v1} (2015).

15. Tafti, F. F., Gibson, Q. D., Kushwaha, S. K., Haldolaarachchige, N. \& Cava, R. J. Resistivity plateau and extreme magnetoresistance in LaSb. \textit{Nat. Phys.} \textbf{12}, 272-277 (2016).

16. Sun, S., Wang, Q., Guo, P.-J., Liu, K. \& Lei, H. Large magnetoresistance in LaBi: origin of field-induced resistivity upturn and plateau in compensated semimetals. \textit{New J. Phys.} \textbf{18}, 082002 (2016).

17. Kumar, N. \textit{et al.} Observation of pseudo-two-dimensional electron transport in the rock salt-type topological semimetal LaBi. \textit{Phys. Rev. B} \textbf{93}, 241106(R) (2016).

18. Kumar, N., Shekhar, C., Klotz, J., Wosnitza, J. \& Felser C. Large valley polarization of electrons in the topological semimetal LaBi. \textit{arXiv:1703.02331}.

19. Guo, P.-J., Yang, H.-C., Zhang, B. J., Liu, K. \& Lu, Z.-Y. Charge compensation in extremely large magnetoresistance materials LaSb and LaBi revealed by first-principles calculations. \textit{Phys. Rev. B} \textbf{93}, 235142 (2016).

20. Zeng, L.-K. \textit{et al.} Compensated Semimetal LaSb with Unsaturated Magnetoresistance. \textit{Phys. Rev. Lett.} \textbf{117}, 127204 (2016).

21. Niu, X. H. \textit{et al.} Presence of exotic electronic surface states in LaBi and LaSb. \textit{Phys. Rev. B} \textbf{94}, 165163 (2016).

22. Nayak, J. \textit{et al.} Multiple Dirac cones at the surface of the topological metal LaBi. \textit{Nat. Comm.} \textbf{8}, 13942 (2017).

23. Lou, R. \textit{et al.} Evidence of topological insulator state in LaBi semimetal. \textit{Phys. Rev. B} \textbf{95}, 115140 (2017).

24. Wu, Y. \textit{et al.} Asymmetric mass acquisition in LaBi: Topological semimetal candidate. \textit{Phys. Rev. B} \textbf{94}, 081108(R) (2016).

25. Tafti, F. F. \textit{et al.} Temperature-field phase diagram of extreme magnetoresistance. \textit{Proc. Natl. Acad. Sci.} \textbf{113}, 3475-3481 (2016).

26. Wang, L.-X., Li, C.-Z., Yu, D.-P. \& Liao, Z.-M. Aharonov-Bohm oscillations in Dirac semimetal Cd$_{3}$As$_{2}$ nanowires. \textit{Nat. Comm.} \textbf{7}, 10769 (2016).

27. Jauregui, L. A., Pettes, M. T., Rokhinson, L. P., Shi, L. \& Chen, Y. P. Magnetic field-induced helical mode and topological transitions in a topological insulator nanoribbon. \textit{Nat. Nanotech.} \textbf{11}, 345–351 (2016).

28. Singha, R., Pariari, A. K., Satpati, B. \& Mandal, P. Large nonsaturating magnetoresistance and signature of nondegenerate Dirac nodes in ZrSiS. \textit{Proc. Natl. Acad. Sci.} \textbf{114(10)}, 2468-2473 (2017).

29. Wang, Y. L. \textit{et al.} Origin of the turn-on temperature behavior in WTe$_{2}$. \textit{Phys. Rev. B} \textbf{92}, 180402(R) (2015).

30. Singha, R., Satpati, B. \& Mandal, P. Low-temperature resistivity plateau and large magnetoresistance in compensated semimetal LaSbTe. \textit{arXiv:1609.09397} (2016).

31. Kopelevich, Y., Pantoja, J. C. M., da Silva, R. R. \& Moehlecke, S. Universal magnetic-field-driven metal-insulator-metal transformations in graphite and bismuth. \textit{Phys. Rev. B} \textbf{73}, 165128 (2006).

32. Wang, Y.-Y., Yu, Q.-H., Guo, P.-J., Liu, K. \& Xia, T.-L. Resistivity plateau and extremely large magnetoresistance in NbAs$_{2}$ and TaAs$_{2}$. \textit{Phys. Rev. B} \textbf{94}, 041103(R) (2016).

33. Ziman, J. M. Electrons and Phonons: The Theory of Transport Phenomena in Solids. \textit{Classics Series, Oxford University Press, New York} (2001).

34. Narayanan, A. \textit{et al.} Linear Magnetoresistance Caused by Mobility Fluctuations in n-Doped Cd$_{3}$As$_{2}$. \textit{Phys. Rev. Lett.} \textbf{114}, 117201 (2015).

35. Murakawa, H. \textit{et al.} Detection of Berry's Phase in a Bulk Rashba Semiconductor. \textit{Science} \textbf{342}, 1490-1493 (2013).

36. Hurd, C. M. The Hall effect in metals and alloys. \textit{Plenum Press, New York} (1972).

37. Luo, Y. \textit{et al.} Hall effect in the extremely large magnetoresistance semimetal WTe$_{2}$. \textit{App. Phys. Lett.} \textbf{107}, 182411 (2015).

38. Hasegawa, A. Fermi surface of LaSb and LaBi. \textit{J. Phy. Soc. Japan} \textbf{54(2)}, 677-684 (1985).

39. Yoshida, M. \textit{et al.} Cyclotron Resonance of LaBi. \textit{J. Phy. Soc. Japan} \textbf{70(7)}, 2078–2081 (2001).

40. Koyama, K., \textit{et al.} Observation of cyclotron resonance in rare-earth monopnictides microwave region. \textit{J. Phy. Chem. Solids} \textbf{63}, 1227-1230 (2002).

41. Yoshida, M., Koyama K., Ochiai, A. \& Motokawa M. Cyclotron resonance in rare-earth monopnictides. \textit{Phys. Rev. B} \textbf{71}, 075102 (2005).

42. Kitazawa, H. \textit{et al.} De Haas-van Alphen effects on La(Sb, Bi) and Ce(Sb, Bi). \textit{J. Mag. Mag. Mater.} \textbf{31-34} 421-422 (1983).

43. Shoenberg, D. Magnetic Oscillations in Metals. \textit{Cambridge Univ. Press} (1984).

44. Hu, J. \textit{et al.} Evidence of Topological Nodal-Line Fermions in ZrSiSe and ZrSiTe. \textit{Phys. Rev. Lett.} \textbf{117}, 016602 (2016).

45. Zhao, Y. \textit{et al.} Anisotropic Fermi Surface and Quantum Limit Transport in High Mobility Three-Dimensional Dirac Semimetal Cd$_{3}$As$_{2}$. \textit{Phys. Rev. X} \textbf{5}, 031037 (2015).

46. Zhang, H., Liu, C.-X. \& Zhang, S.-C. Spin-Orbital Texture in Topological Insulators. \textit{Phys. Rev. Lett.} \textbf{111}, 066801 (2013).

47. Hsieh, D. \textit{et al.} A tunable topological insulator in the spin helical Dirac transport regime. \textit{Nature} \textbf{460}, 1101-1105 (2009).

48. Zhao, L. \textit{et al.} Singular robust room-temperature spin response from topological Dirac fermions. \textit{Nat. Mater.} \textbf{13}, 580-585 (2014).

49. Buga, S. G. \textit{et al.} Superconductivity in bulk polycrystalline metastable phases of Sb$_{2}$Te$_{3}$ and Bi$_{2}$Te$_{3}$ quenched after high-pressure-high-temperature treatment. \textit{Chem. Phys. Lett.} \textbf{631-632}, 97-102 (2015).

50. Dutta, P., Pariari, A. \& Mandal, P. Prominent metallic surface conduction and the singular magnetic response of topological Dirac fermion in three-dimensional topological insulator Bi$_{1.5}$Sb$_{0.5}$Te$_{1.7}$Se$_{1.3}$. \textit{arXiv:1608.01203v2} (2016).

51. Pariari, A. \& Mandal, P. Coexistence of topological Dirac fermions in the surface and three-dimensional Dirac cone state in the bulk of ZrTe$_{5}$ single crystal. \textit{Sci. Rep.} \textbf{7}, 40327 (2017).

52. Canfield, P.C. \& Fisk, Z. Growth of single crystals from metallic fluxes. \textit{Phil. Mag. B} \textbf{65(6)}, 1117-1123 (1992).\\

\textbf{Acknowledgements}

We acknowledge and thank A. Pariari, P. Dutta, S. Roy and A. Pal for their help during measurements and fruitful discussions.

\textbf{Author Contributions}

P.M. designed research. R.S. prepared the crystals and performed experiments. B.S performed HRTEM measurement. R.S and P.M analyzed data and wrote the paper. P.M. supervised the project.

\textbf{Competing financial interests:} The authors declare no competing financial interests.\\

\end{document}